\documentclass[aps,superscriptaddress,twocolumn,showpacs,preprintnumbers,amsmath,amssymb]{revtex4}
\usepackage[cp1251]{inputenc}
\usepackage[english]{babel}
\usepackage{amssymb,amsmath}
\usepackage{amstext} 
\usepackage[mathscr]{eucal}
\usepackage[pdftex]{graphicx}
\usepackage{longtable}
\usepackage[unicode, colorlinks=false,linkcolor=blue, pdfborder={0 0 0}]{hyperref} 
\begin{document}
\title{To Cavity EM-Field Quantization} 
\author{A.Dovlatova}
\affiliation{M.V.Lomonosov Moscow State University, Moscow, 119899}
\author{D.Yearchuck}
\affiliation{Minsk State Higher College, Uborevich Str.77, Minsk, 220096, RB; yearchuck@gmail.com}
\author{Y.Yerchak}
\affiliation{Belarusian State University, Nezavisimosti Ave.4, Minsk, 220030, RB}
\author{A.Alexandrov}
\affiliation{M.V.Lomonosov Moscow State University, Moscow, 119899}
\date{\today}
\begin{abstract}Cavity QED canonical quantization theory is developed, taking into consideration the dual symmetry of Maxwell equations. The expression for the charge quantum is established for the first time.
\end{abstract}
\pacs{78.20.Bh, 75.10.Pq, 11.30.-j, 42.50.Ct, 76.50.+g}
\maketitle
Cavity QED theory  is substantial for many practical applications, for instance  for
implementing of quantum computation, and is studied extensively.
In the theory of cavity QED, the Jaynes-
Cummings  model (JCM)  for the one qubit 
case  is recognized to be the simplest
and at the same time the most effective model for the interaction between quantized electromagnetic (EM) field 
 and the  matter, which can be solved exactly. JCM predicts among many interesting features, the occurrence of collapses and revivals of coherence in the dynamics
of a two-level atom in interaction with a single mode of the quantized EM-field, squeezing, antibunching, chaos, etc.

It seems to be, that the spectrum of quantization consequences,  established at present, is far from complete even for simple models.  The fact is that, that used at present canonical quantization procedure, which was proposed for the first time still at the earliest stage of quantum physics \cite{Born}, is needing in its development.

The fact is, the existing at present quantization procedure, being to be based on Maxwell equations, does not take nevertheless into consideretion the dual symmetry of given equations.  

The aim of given work is to develop the canonical quantization procedure for cavity QED.

Suppose EM-field in volume rectangular cavity. Suppose also, that the field polarization is linear in z-direction. Then the vector of electrical component can be represented in the form 
\begin{equation}
\label{eq1}
\vec{E}^{[1]}(\vec{r},t) = 
E_x(z,t) \vec{e}_x = \left[\sum_{\alpha=1}^{\infty}A_{\alpha}q_{\alpha}(t)\sin(k_{\alpha}z)\right]\vec{e}_x ,
\end{equation}
where $q_{\alpha}(t)$ is amplitude of $\alpha$-th normal mode of the cavity, $\alpha \in N$, $k_{\alpha} = \alpha\pi/L$, $A_{\alpha}=\sqrt{2 \omega_{\alpha}^2m_{\alpha}/(V\epsilon_0)}$, $\omega_{\alpha} = \alpha\pi c/L$, $L$ is cavity length along z-axis, $V$ is cavity volume, $m_{\alpha}$ is parameter, which is introduced to obtain the analogy with mechanical harmonic oscillator. Let us analyse the solution of Maxwell equations for EM-field in a cavity in comparison with known solution from the literature to pay the attention to some mathematical details. Using the equation
\begin{equation}
\label{eq2}
\epsilon_0 \frac{\partial\vec{E}(z,t)}{\partial t } = \left[\nabla\times\vec{H}(z,t)\right],
\end{equation}
 we obtain  the expression for magnetic field
 \begin{equation}
\label{eq3}
\vec{H}(\vec{r},t) =  \left[\sum_{\alpha=1}^{\infty}A_{\alpha}\frac{\epsilon_0}{k_{\alpha}}\frac{dq_{\alpha}(t)}{dt}\cos(k_{\alpha}z) + f_{\alpha}(t)\right]\vec{e}_y,
\end{equation}   
where  $\{f_{\alpha}(t)\}$, $ \alpha \in N $, is the set of arbitrary  functions of the time.
The partial solution, in which the functions $\{f_{\alpha}(t)\}$ are identically zero, is always used in all the EM-field literature. However even in given case it is evident, that the Maxwellian field is complex field. Really 
 using the equation
\begin{equation}
\label{eq4}
\left[ \nabla\times\vec{E}\right] = -\frac{\partial \vec{B}}{\partial t} = -\mu_0 \frac{\partial \vec{H}}{\partial t}
\end{equation}
it is easily to find the class of field functions $\{q_{\alpha}(t)\}$. They will satisfy to  differential equations
\begin{equation}
\label{eq5}
\frac{d^2q_{\alpha}(t)}{dt^2} + \frac{k_{\alpha}^2}{\mu_0\epsilon_0} q_{\alpha}(t)=0, \alpha \in N.
\end{equation}
Consequently,  we have
\begin{equation}
\label{eq6}
q_{\alpha}(t) = C_{1\alpha} e^{i\omega_{\alpha}t} + C_{2\alpha} e^{-i\omega_{\alpha}t}, \alpha \in N,
\end{equation}
where  $C_{1\alpha}, C_{2\alpha}, \alpha \in N$ are arbitrary constants.
Thus, real-valued free Maxwell field equations result in well known in the theory of differential equations  
situation - the solutions are complex-valued functions. It means, that generally the field functions for free Maxwellian field in the cavity produce complex space. So the  known conception, that EM-field is real-valued, has to be corrected. 

Further,
there is the second physically substantial solution of Maxwell equations. Really from general expression (\ref{eq3}) for the  field $\vec{H}(\vec{r},t)$ 
it  is easily to obtain differential equations  for $\{f_{\alpha}(t)\}$, $ \alpha \in N $
\begin{equation}
\label{eq7}
\begin{split}
&\frac{d f_{\alpha}(t)}{dt} + A_{\alpha}\frac{\epsilon_0}{k_{\alpha}}\frac{\partial^2q_{\alpha}(t)}{\partial t^2}\cos(k_{\alpha}z) \\
&- \frac {1}{\mu_0} A_{\alpha}k_{\alpha}q_{\alpha}(t)cos(k_{\alpha}z) = 0.
\end{split}
\end{equation}
The  formal solution of given equations 
in general case is
\begin{equation}
\label{eq8}
f_{\alpha}(t) =  A_{\alpha} \cos(k_{\alpha}z)\left[\frac{k_{\alpha}}{\mu_0} \int\limits _{0}^{t} q_{\alpha}(\tau)d\tau -\frac{dq_{\alpha}(t)}{dt}\frac{\epsilon_0}{k_{\alpha}}\right]
\end{equation}
Therefore, we have the second  solution  of Maxwell equations  in the form
\begin{equation}
\label{eq9}
\vec{H}^{[2]}(\vec{r},t) = \frac{1}{\mu_0}\left\{\sum_{\alpha=1}^{\infty}k_{\alpha}A_{\alpha} \cos(k_{\alpha}z) q_{\alpha}'(t)\right\}\vec{e}_y,
\end{equation}
\begin{equation}
\label{eq10}
\vec{E}^{[2]}(\vec{r},t) = \left\{\sum_{\alpha=1}^{\infty}A_{\alpha}\frac{dq_{\alpha}'(t)}{dt}\sin(k_{\alpha}z)\right\}\vec{e}_x,
\end{equation}
where
\begin{equation}
\label{eq11}
q_{\alpha}'(t) = \int\limits _{0}^{t} q_{\alpha}(\tau)d\tau.
\end{equation}

The field Hamiltonian $\mathcal{H}^{[1]}(t)$, corresponding to the first  partial solution, in canonical form is
\begin{equation}
\label{eq12}
\begin{split}
&\mathcal{H}^{[1]}(t) = \frac{1}{2}\iiint\limits_{(V)}\left[\epsilon_0E_x^2(z,t)+\mu_0H_y^2(z,t)\right]dxdydz\\
&= \frac{1}{2}\sum_{\alpha=1}^{\infty}\left[m_{\alpha}\nu_{\alpha}^2q_{\alpha}^2(t) + \frac{p_{\alpha}^2(t)}{m_{\alpha}} \right], p_{\alpha} = m_{\alpha} \frac{dq_{\alpha}(t)}{dt}.
\end{split}
\end{equation}

The Hamiltonian $\mathcal{H}^{[2]}(t)$, corresponding to the second solution, can also be represented in canonical form
\begin{equation}
\label{eq14}
\mathcal{H}^{[2]}(t) = 
\frac{1}{2}\sum_{\alpha=1}^{\infty}\left[m_{\alpha}\nu_{\alpha}^2q{''}_{\alpha}^2(t) + \frac{p{''}_{\alpha}^2(t)}{m_{\alpha}} \right],
\end{equation}
where
\begin{equation}
\label{eq15}
\begin{split}
q{''}_{\alpha}(t) = \nu_{\alpha}q{'}_{\alpha}(t),  
p{''}_{\alpha}(t) = m_{\alpha}\nu_{\alpha}\frac{dq_{\alpha}'(t)}{dt}
\end{split}
\end{equation}
Here we wish to pay attention, that the analogy with mechanical oscillator is the only partial, since in mechanics the canonical variables are the components of the vectors of the same parity (of polar vectors for harmonic oscillator). In EM-field theory vector-functions with different parities correspond to  "coordinates" $q{}_{\alpha}(t), q{''}_{\alpha}(t)$ and to "impulses" $p_{\alpha}(t), p{''}_{\alpha}(t)$.   If canonical "coordinates" are determined by polar vector-functions, then canonical "impulses" are axial vector-functions. We see, that for two partial independent solutions the roles of electric and magnetic vector-functions trade places. Consequently they both can be polar and axial vector-functions, which follows immediately from above represented  partial solutions of Maxwell equations. In correspondence with conclusion on complex nature of Maxwellian field and 
in accordance with definition of complex quantities we can represent both partial solutions in the form
\begin{equation}
\label{eq16}
(\vec{E}(\vec{r},t), \vec{E}^{[2]}(\vec{r},t)) \rightarrow \vec{E}(\vec{r},t) +  i \vec{E}^{[2]}(\vec{r},t) = \vec{E}_c(\vec{r},t).
\end{equation}
and
\begin{equation}
\label{eq17}
(\vec{H}^{[2]}(\vec{r},t), \vec{H}(\vec{r},t)) \rightarrow \vec{H}^{[2]}(\vec{r},t) +  i \vec{H}(\vec{r},t) = \vec{H}_c(\vec{r},t)
\end{equation}
Given solution possess by dual symmetry. It means that both electric and magnetic EM-field vector-functions can be even and uneven under spatial inversion transformations. On the other hand either of the two given kinds in accordance with (\ref{eq6}) consists of the parts, which are even and uneven under time reversal transformations. We see on the example considered,  that  free EM-field is  4-fold degenerated, that is consist of four components with different spatial inversion and time reversal parity.

 At the same time the solution of Maxwell equations in the form (\ref{eq16}) and (\ref{eq17}) means, that we come to complex form of Maxwell equations by a natural way. 

It represents the interest to calculate the 4-currents for given task. It is evident, that  
\begin{equation}
\label{eq18}
{j_\mu} = {j_\mu}^{(1)} + i {j_\mu}^{(2)},
\end{equation}
where
${j_\mu}^{(1)}$ is well known quantity, which is determined by
\begin{equation}
\label{eq19}
\begin{split}
&{j_\mu}^{(1)}(x) = -\frac{i e}{\hbar c}\sum_{\alpha = 1}^{\infty}\sum_{j=1}^{2}\left[ \frac{\partial{L}}{\partial(\partial_{\mu}u^{j}_\alpha(x))} u^{j}_{\alpha}(x)\right] \\
&+ \frac{i e}{\hbar c}\sum_{\alpha = 1}^{\infty}\sum_{j=1}^{2}\left[ \frac{\partial{L}}{\partial(\partial_{\mu}u^{j}_\alpha(x))^*} u^{*j}_{\alpha}(x)\right].
\end{split}
\end{equation}
To determine the current ${j_\mu}^{(2)}$ we have to take into consideration, that gauge symmetry group of EM-field is two-parametric group $\Gamma(\alpha,\beta) = U_{1}(\alpha) \otimes \mathfrak R(\beta)$, where $\mathfrak R(\beta)$ is abelian multiplicative group of real numbers (excluding zero). It leads in fact to existence of complex current including complex charge for EM-field. Physically the presence of additional gauge symmetry above indicated follows from known invariance of Maxwell equations under scale transformations. Then, it can be shown, using N\"{o}ther theorem, that 
the current ${j_\mu}^{(2)}$ is
\begin{equation}
\label{eq20}
\begin{split}
&{j_\mu}^{(2)}(x) = -\frac{e}{\hbar c}\sum_{\alpha = 1}^{\infty}\sum_{j=1}^{2}\left[ \frac{\partial{L}}{\partial(\partial_{\mu}u^{j}_\alpha(x))} u^{j}_{\alpha}(x)\right] \\
&- \frac{e}{\hbar c}\sum_{\alpha = 1}^{\infty}\sum_{j=1}^{2}\left[ \frac{\partial{L}}{\partial(\partial_{\mu}u^{j}_\alpha(x))^*} u^{*j}_{\alpha}(x)\right].
\end{split}
\end{equation}
 Lagrangian $L(x)$ can be represented in the following form
\begin{equation}
\label{eq21}
\begin{split}
&L(x) = \frac{1}{2}\sum_{j=1}^{2}\sum_{\mu=1}^{4}\sum_{\alpha=1}^{\infty}( c^2)  \frac{\partial u^{j}_{\alpha}(x)}{\partial x_\mu}\left[\frac{\partial u^{j}_{\alpha}(x)}{\partial x_\mu}\right]^{*} \\
&- \frac{1}{2}\sum_{j=1}^{2}\sum_{\mu=1}^{4}\sum_{\alpha=1}^{\infty} \omega_{\alpha}^2 u^{j}_{\alpha}(x) u^{*j}_{\alpha}(x).
\end{split}
\end{equation}
Let us choose the set of time dependent components $ \{ q_{\alpha}(t) \}$ of field vector-functions in the form
\begin{equation}
\label{eq22}
 \{ q_{\alpha}(t) \} = \{ e^{i \omega_{\alpha} t} \},
\end{equation}
then we will have for the components of 4-vectors  ${j_\mu}^{(1)}$ and  ${j_\mu}^{(2)}$
\begin{equation}
\label{eq23}
\begin{split}
&{j_\mu}^{(1)} = 0,  \mu = \overline {1,4}, \\
&{j_\mu}^{(2)} = 0, \mu = \overline {1,3}, {j_4}^{(2)} = -\frac{2ec}{\hbar }\sum_{\alpha = 1}^{\infty} A^{2}_{\alpha} \omega_{\alpha}.
\end{split}
 \end{equation}
CGS system is used, that is 
$\vec B(x) = \vec H(x), \vec D(x) = \vec E(x)$.
It is easily to make sure, that continuity equation
\begin{equation}
\label{eq25}
\frac{\partial j_\mu}{\partial x_\mu} = 0 
 \end{equation}
is fulfilled well for the case considered. It means, that imaginary part of charge, which is nonzero, is really conserving quantity. 
We use further the standard procedure of field quantization. So for the first partial solution we have
\begin{equation}
\label{eq26}
\begin{split}
&\left[\hat {p}_{\alpha}(t) , \hat {q}_{\beta}(t)\right] = i\hbar\delta_{{\alpha}\beta}\\
&\left[\hat {q}_{\alpha}(t) , \hat {q}_{\beta}(t)\right] = \left[\hat {p}_{\alpha}(t) , \hat {p}_{\beta}(t)\right] = 0,
\end{split}
\end{equation}
where
$\alpha, \beta \in N$.
Introducing the operators $\hat{a}_{\alpha}(t)$ and $ \hat{a}^{+}_{\alpha}(t)$
\begin{equation}
\label{eq27}
\begin{split}
&\hat{a}_{\alpha}(t) = \frac{1}{ \sqrt{ 2 \hbar  m_{\alpha} \omega_{\alpha}}} \left[ m_{\alpha} \omega_{\alpha}\hat {q}_{\alpha}(t) + i \hat {p}_{\alpha}(t)\right]\\
&\hat{a}^{+}_{\alpha}(t) = \frac{1}{ \sqrt{ 2 \hbar  m_{\alpha} \omega_{\alpha}}} \left[ m_{\alpha} \omega_{\alpha}\hat {q}_{\alpha}(t) - i \hat {p}_{\alpha}(t)\right],
\end{split}
\end{equation}
we have for the operators of canonical variables
\begin{equation}
\label{eq28}
\begin{split}
&\hat {q}_{\alpha}(t) = \sqrt{\frac{\hbar}{2 m_{\alpha} \omega_{\alpha}}} \left[\hat{a}^{+}_{\alpha}(t) + \hat{a}_{\alpha}(t)\right]\\
&\hat {p}_{\alpha}(t) = i \sqrt{\frac{\hbar m_{\alpha} \omega_{\alpha}}{2}} \left[\hat{a}^{+}_{\alpha}(t) - \hat{a}_{\alpha}(t)\right]. 
\end{split}
\end{equation}
Then field function operators are
\begin{equation}
\label{eq29}
\hat{\vec{E}}^{[1]}(\vec{r},t) = \{\sum_{\alpha=1}^{\infty} \sqrt{\frac{\hbar \omega_{\alpha}}{V\epsilon_0}} \left[\hat{a}^{+}_{\alpha}(t) + \hat{a}_{\alpha}(t)\right] sin(k_{\alpha} z)\} \vec{e}_x,
\end{equation}

\begin{equation}
\label{eq30}
\hat{\vec{H}}^{[1]}(\vec{r},t) = i \{\sum_{\alpha=1}^{\infty} \sqrt{\frac{\hbar \omega_{\alpha}}{V\mu_0}} \left[\hat{a}^{+}_{\alpha}(t) - \hat{a}_{\alpha}(t)\right] cos(k_{\alpha} z)\} \vec{e}_y,
\end{equation} 
For the second partial solution, corresponding to Hamiltonian  $\mathcal{H}^{[2]}(t)$ we have
\begin{equation}
\label{eq31}
\begin{split}
&\left[\hat{p}{''}_{\alpha}(t) , \hat {q}{''}_{\beta}(t)\right] = i\hbar\delta_{{\alpha}\beta}\\
&\left[\hat {q}{''}_{\alpha}(t) , \hat {q}{''}_{\beta}(t)\right] = \left[\hat {p}{''}_{\alpha}(t) , \hat {p}{''}_{\beta}(t)\right] = 0,
\end{split}
\end{equation}
$\alpha, \beta \in N$.
The operators $\hat{a}{''}_{\alpha}(t)$, $\hat{a}{''}^{+}_{\alpha}(t)$ are introduced analogously
\begin{equation}
\label{eq32}
\begin{split}
&\hat{a}{''}_{\alpha}(t) = \frac{1}{ \sqrt{ 2 \hbar  m_{\alpha} \omega_{\alpha}}} \left[ m_{\alpha} \omega_{\alpha}\hat {q}{''}_{\alpha}(t) + i \hat {p}{''}_{\alpha}(t)\right]\\
&\hat{a}{''}^{+}_{\alpha}(t) = \frac{1}{ \sqrt{ 2 \hbar  m_{\alpha} \omega_{\alpha}}} \left[ m_{\alpha} \omega_{\alpha}\hat {q}{''}_{\alpha}(t) - i \hat {p}{''}_{\alpha}(t)\right]
\end{split}
\end{equation}
Relationships for canonical variables are
\begin{equation}
\label{eq33}
\begin{split}
&\hat {q}{''}_{\alpha}(t) = \sqrt{\frac{\hbar}{2 m_{\alpha} \omega_{\alpha}}} \left[\hat{a}{''}^{+}_{\alpha}(t) + \hat{a}{''}_{\alpha}(t)\right]\\
&\hat {p}{''}_{\alpha}(t) = i \sqrt{\frac{\hbar m_{\alpha} \omega_{\alpha}}{2}} \left[\hat{a}{''}^{+}_{\alpha}(t) - \hat{a}{''}_{\alpha}(t)\right] 
\end{split}
\end{equation}
 For the field function operators we obtain
\begin{equation}
\label{eq34}
\begin{split}
&\hat{\vec{E}}^{[2]}(\vec{r},t) = \\
&i \{\sum_{\alpha=1}^{\infty} \sqrt{\frac{\hbar \omega_{\alpha}}{V\epsilon_0}} \left[\hat{a}{''}^{+}_{\alpha}(t) - \hat{a}{''}_{\alpha}(t)\right] sin(k_{\alpha} z)\} \vec{e}_x,
\end{split}
\end{equation}
\begin{equation}
\label{eq35}
\begin{split}
&\hat{\vec{H}}^{[2]}(\vec{r},t) = \\
&\{\sum_{\alpha=1}^{\infty} \sqrt{\frac{\hbar \omega_{\alpha}}{V\mu_0}} (-1) \left[\hat{a}{''}^{+}_{\alpha}(t) + \hat{a}{''}_{\alpha}(t)\right] cos(k_{\alpha} z)\} \vec{e}_y.
\end{split}
\end{equation}
Representing field operators  in the form (\ref{eq16}) and (\ref{eq17})  we  will have
 \begin{equation}
\begin{split}
&\hat{\vec{E}}(\vec{r},t) =  \{\sum_{\alpha=1}^{\infty} \sqrt{\frac{\hbar \omega_{\alpha}}{V\epsilon_0}} \{\left[\hat{a}^{+}_{\alpha}(t) + \hat{a}_{\alpha}(t)\right]\\
& + \left[\hat{a}{''}_{\alpha}(t) - \hat{a}{''}^{+}_{\alpha}(t)\right]\} sin(k_{\alpha} z)\} \vec{e}_x,
\end{split}
\end{equation}
and
\begin{equation}
\begin{split}
&\hat{\vec{H}}(\vec{r},t) =  \{\sum_{\alpha=1}^{\infty} \sqrt{\frac{\hbar \omega_{\alpha}}{V\mu_0}}\{ \left[\hat{a}^{+}_{\alpha}(t) - \hat{a}_{\alpha}(t)\right] \\
&- \left[\hat{a}{''}_{\alpha}(t) + \hat{a}{''}^{+}_{\alpha}(t)\right] \} cos(k_{\alpha} z) \} \vec{e}_y,
\end{split}
\end{equation}
It can be shown, that
\begin{equation}
\begin{split}
&\hat{j}^{(1)}_4(x) = \frac{ie}{\hbar^2}\sum_{\alpha=1}^{\infty}K_{1 \alpha}\frac{\hbar \omega_{\alpha}}{V\epsilon_0}\sin^2k_{\alpha}z  \left[\hat{a}^{+}_{\alpha}(x_4), \hat{a}_{\alpha}(x_4)\right]\\ 
&+ \frac{ie}{\hbar^2}\sum_{\alpha=1}^{\infty}K_{1 \alpha} \frac{\hbar \omega_{\alpha}}{V\epsilon_0}\sin^2k_{\alpha}z \{\left[\hat{a}^{+}_{\alpha}(x_4)\right]^2 -\left[\hat{a}_{\alpha}(x_4)\right]^2\}\\
&+ \frac{ie}{\hbar^2}\sum_{\alpha=1}^{\infty} K_{1 \alpha}\frac{\hbar \omega_{\alpha}}{V\epsilon_0}\sin^2k_{\alpha}z \left[\hat{a}^{''+}_{\alpha}(x_4), \hat{a}^{''}_{\alpha}(x_4)\right] \\
&- \frac{ie}{\hbar^2}\sum_{\alpha=1}^{\infty} K_{1 \alpha}\frac{\hbar \omega_{\alpha}}{V\epsilon_0}\sin^2k_{\alpha}z \{\left[\hat{a}^{''+}_{\alpha}(x_4)\right]^2  + \left[\hat{a}^{''}_{\alpha}(x_4)\right]^2\} \\
& -  \frac{ie}{\hbar^2}\sum_{\alpha=1}^{\infty} K_{2 \alpha}\frac{\hbar \omega_{\alpha}}{V \mu_0} \cos^2k_{\alpha}z  \left[\hat{a}_{\alpha}(x_4), \hat{a}^{+}_{\alpha}(x_4)\right]\\
&+ \frac{ie}{\hbar^2}\sum_{\alpha=1}^{\infty} K_{2 \alpha}\frac{\hbar \omega_{\alpha}}{V \mu_0} \cos^2k_{\alpha}z \{\left[\hat{a}^{+}_{\alpha}(x_4)\right]^2 -\left[\hat{a}_{\alpha}(x_4)\right]^2\}\\
&+ \frac{ie}{\hbar^2}\sum_{\alpha=1}^{\infty} K_{2 \alpha}\frac{\hbar \omega_{\alpha}}{V \mu_0} \cos^2k_{\alpha}z \left[\hat{a}^{''+}_{\alpha}(x_4), \hat{a}^{''}_{\alpha}(x_4)\right]\\
&- \frac{ie}{\hbar^2}\sum_{\alpha=1}^{\infty} K_{2 \alpha} \frac{\hbar \omega_{\alpha}}{V \mu_0} \cos^2k_{\alpha}z \{\left[\hat{a}^{''+}_{\alpha}(x_4)\right]^2  + \left[\hat{a}^{''}_{\alpha}(x_4)\right]^2\},  
\end{split}
\end{equation}
where $\hat{j}^{(1)}_4(x)$ is operator of electric charge density, $K_{1\alpha}$, $K_{2\alpha}$, $\alpha \in N$ are the constants, which can be determined  from initial or boundary conditions. 
The expression for $\hat{j}^{(2)}_4(x)$ is
\begin{equation}
\begin{split}
&\hat{j}^{(2)}_4(x) = \frac{ie}{\hbar^2}\sum_{\alpha=1}^{\infty}K_{1\alpha}\frac{\hbar \omega_{\alpha}}{V\epsilon_0}\sin^{2}k_{\alpha}z  \{\hat{a}^{+}_{\alpha}(x_4), \hat{a}^{''}_{\alpha}(x_4)\} \\
&+ \frac{ie}{\hbar^2}\sum_{\alpha=1}^{\infty}\frac{K_{1\alpha}\hbar \omega_{\alpha}}{V\epsilon_0}\sin^{2}k_{\alpha}z \{\hat{a}_{\alpha}(x_4), \hat{a}^{''+}_{\alpha}(x_4)\}\\
&+ \frac{ie}{\hbar^2}\sum_{\alpha=1}^{\infty} K_{2\alpha} \frac{\hbar \omega_{\alpha}}{V \mu_0} \cos^{2}k_{\alpha}z \{\hat{a}^{+}_{\alpha}(x_4), \hat{a}^{''}_{\alpha}(x_4)\} \\
&+ \frac{ie}{\hbar^2}\sum_{\alpha=1}^{\infty} K_{2\alpha} \frac{\hbar \omega_{\alpha}}{V \mu_0} \cos^{2}k_{\alpha}z \{\hat{a}_{\alpha}(x_4), \hat{a}^{''+}_{\alpha}(x_4)\},\\  
&\alpha \in N,
\end{split}
\end{equation}
where $\hat{j}^{(2)}_4(x)$ is operator of magnetic charge density. Observable value of the electric charge density for the field state with $n$ photons in $\alpha$ mode and with uneven  spatial parity then is
\begin{equation}
\langle n_{\alpha}|\hat{j}^{(1)}_4(x) |n_{\alpha}\rangle = - \frac{ie\omega_{\alpha}}{\hbar V }\left[\frac{K_{1 \alpha}}{\epsilon_0}\sin^{2}k_{\alpha}z + \frac{K_{2 \alpha}}{\mu_0} \cos^{2}k_{\alpha}z \right]
\end{equation}
Conserving quantity, that is quantized electric charge is
\begin{equation}
q_{\alpha} = \frac{e\omega_{\alpha}}{2 \hbar} \left[\frac{K_{1 \alpha}}{\epsilon_0} + \frac{K_{2 \alpha}}{\mu_0}\right] \equiv g \omega_{\alpha},
\end{equation}
where
\begin{equation}
g = \frac{e}{2 \hbar} \left[\frac{K_{1 \alpha}}{\epsilon_0} + \frac{K_{2 \alpha}}{\mu_0} \right]
\end{equation}
It is easily to show that observable value of the magnetic charge density for the field state with $n$ photons in $\alpha$ mode and with uneven  spatial parity is equal to zero. We see, that classical and quantum consideration lead to different results. We see also that the expression for charge quantum is simple and it is similar to the expression for the energy quantum. It is evident that quantization of electric charge and magnetic charge in the cavity QED are realized independenly, in distinction from Dirac consideration \cite{Dirac} for free EM-field. It is understandable also that the nonzeroth value of magnetic charge quantum will always substantially exceed the value of electric charge quantum, however their ratio in general case can be dependent on initial or boundary conditions in the cavity.

Recently the phenomena of ferroelectric spin wave resonance (FE SWR) and antiferroelectric spin wave resonance (AF SWR) were discovered experimentally and reported in \cite{Yearchuck_Yerchak} and \cite{Yearchuck_PL}. The experimental results were explained in the frame of the model reported in \cite{Yearchuck_Doklady}, in which the existence of the EM-field component with axial electric field vector-function was required to obtain the agreement with earlier known experimentally detected optical analogues of magnetic resonance transition phenomena. The ferromagnetic spin wave resonance (FM SWR) was also observed earlier strictly in the same sample, that is on the same carbon chains, and was reported  in   \cite{Ertchak_J_Physics_Condensed_Matter}.
The values of splitting parameters $\mathfrak{A}^E$ and $\mathfrak{A}^H$ in FE SWR and FM SWR allow to find 
 the ratio $J_{E }/J_{H}$ of exchange 
constants.  The range of the ratio $J_{E }/J_{H}$ was $(1.2 - 1.6)10^{4}$. Given result seems to be direct proof,  that the function, which is invariant under gauge transformations is two component, that is complex-valued function. In other words, the complex charge corresponds 
to presence of exchange interacting solitons with two independent exchange constants. We can evaluate the ratio of imagine $e_{H} \equiv g$ to real $e_{E}\equiv e $  components of complex charge taking into account the relationship between exchange interaction and charge. It is $\frac{g}{e} \sim \sqrt{J_{E }/J_{H}} \approx (1.1 - 1.3)10^{2}$.  It  seen, that given result agrees well with Dirac relationship \cite{Dirac} $g \simeq 68.5 e n$, where $n = \pm{1},\pm{2},...,$ at $n = 2$. It is substantial that magnetic charge carriers are space extended,  $\sim 20$ interatomic carbon chain units, objects - spin-Peierls solitons, which produce the superlattice in carbynoid samples. They differs naturally from point electric charge carriers - electrons and from hypothetic Dirac point monopoles.

\end{document}